# Single-molecule optical absorption imaging by nanomechanical photothermal sensing at room temperature

Miao-Hsuan Chien[1], Mario Brameshuber[2], Gerhard J. Schütz[2], and Silvan Schmid[1]

[1] Institute of Sensor and Actuator Systems, TU Wien, Gusshausstrasse 27-29, 1040 Vienna, Austria.
[2] Institute of Applied Physics, TU Wien, Wiedner Hauptstrasse 8-10, 1040 Vienna, Austria.

**Absorption microscopy is a powerful technique, enabling the detection of single non-fluorescent molecules at room temperature. So far, the molecular absorption has been probed optically via the attenuation of a probing laser. The sensitivity of optical probing is not only restricted by background scattering, but it is fundamentally limited by laser shot noise. Here, we present nanomechanical photothermal microscopy, which overcomes the scattering and shot noise limit by detecting the sample absorption directly with a temperature sensitive substrate. We use nanomechanical silicon nitride drums, whose resonant frequency detunes with local heating. Individual Au nanoparticles with diameters from 10 nm – 200 nm and single molecules (Atto 633) are scanned with a 305 µW heating laser with a peak irradiance of 330 µW/µm$^2$. Using stress-optimized drums, we achieve a sensitivity of 45 fW/Hz$^{1/2}$, which results in a signal-to-noise ratio of >60 for a single molecule. Our method has important consequences for a wide range of applications, such as imaging, absorption analysis and spectrochemical analysis of non-fluorescent samples.**

Introduction

Optical single molecule detection techniques have become indispensable tools in the scientific community over the past decades. Not only does the high sensitivity enable trace analysis of minute samples, but the analysis of single molecules also provides insight into the individual specific properties, which may differ from the statistical average behaviour of bulk samples. Optical single molecule detection has highly relied on fluorescence microscopy due to its low background and high sensitivity. However, fluorescent emitting states of fluorophores can be easily quenched or destroyed by photochemical reactions. Furthermore, fluorescent labeling has shown to possibly weaken the intermolecular electrostatic interactions and thus change the system dynamics.[1] Label-free optical single molecule detection is thus of fundamental interests for many fields, ranging from tracking and quantification of components in cell biology[2,3] such as viruses, proteins and antibodies, applications in high-accuracy medical diagnostics,[4–8] to environmental monitoring.[9,10] Moreover, the limitations given by photochemical instabilities of fluorescent markers, such as photobleaching and blinking, can also be bypassed.

Optical absorption scales linearly with the volume of the absorber, compared to the less favorable square scaling with the sample volume of optical scattering. This results in larger cross sections for absorption compared to scattering for nanoparticles smaller than ~100 nm.[11] This makes absorption techniques highly effective for detection and imaging in the nm-regime. As a result, absorption-based techniques, such as photothermal microscopy,[12–15] scanning interferometric imaging,[16,17] direct absorption,[18] whispering-gallery mode resonator sensing,[19] and stimulated emission microscopy[20] have been widely developed and investigated with single-nanoparticle sensitivity. In particular, single-molecule sensitivity at room temperature has been achieved by several absorption-based techniques, such as direct optical absorption microscopy,[21] ground-state depletion microscopy,[22] and photothermal contrast.[23,24] However, the detection of the minute relative light attenuation of the order of $10^{-7} – 10^{-6}$ caused by the photon absorption of a single molecule is challenging. The sensitivity is not only limited by scattering due to variations of refractive index in the sample, but more importantly, it is fundamentally limited by shot noise of the probing laser. This results in shot noise limited sensitivities of the order of a few pW/Hz$^{1/2}$.[21,22] For detection of single molecules via their photothermal heating, a sensitivity of a few nW/Hz$^{1/2}$ has been reached.[23,24]

In contrast to optical probing, it has been shown that the absorption of single gold[25] and polymer[26] nanoparticles can be directly detected via the photothermally-induced frequency detuning of nanomechanical string resonators. In essence, the sample substrate acts as the thermometer, which detects the photothermal heating of a local sample such as nanoparticles and molecules. However, the effective scanning region on a slender string resonator is limited to a few micrometers given by its lateral dimensions. Furthermore, the thin nanomechanical structures are fragile, which impedes sampling of liquids and general handling.

Here, we introduce nanomechanical photothermal microscopy, based on drum resonators, as a novel optical platform for single nanoparticle and single molecule imaging and analysis. Our work is based on silicon nitride nanodrums with a typical thickness of 50 nm, which are commonly used as windows for transmission electron microscopy or x-ray diffraction analysis due to their low and featureless background absorption for electrons and photons in the visible to near-infrared regime. Silicon nitride drums enable robust and easy to use sample substrates, which withstand liquid handling and sampling, including pipetting and spin-coating, etc. Furthermore, such nanomechanical silicon nitride drums are available commercially.

The working principle of our method is depicted in Figure 1(a). When the scanning probe laser hits a sample, the optical absorption of the sample causes local heating and a heat flux into the drum resonator. The corresponding thermal expansion reduces the stress of the drum, which causes a detectable detuning in the mechanical resonance frequency. The scanning of the drum was done by a commercial laser-Doppler vibrometer (LDV) with a 633 nm read-out laser and a long working-distance 50× objective. The LDV laser functions simultaneously as pump and probe laser for the excitation of samples and the real-time read-out of the drum vibration, respectively. This bypasses the complexity of typical optical alignments. In order to cancel heat dissipation via the surrounding gas, and thus enhance the sensitivity, the measurements are performed in vacuum below $10^{-4}$ mbar. The drum vibration was driven with a piezoelectric actuator.

**Imaging of single 10 nm Au nanoparticles**

First, 10 nm gold nanoparticles (AuNPs) where imaged using a laser power of 305 µW. The mechanical resonance frequency of the silicon nitride drum was tracked using a phase-locked loop (PLL) during scanning with constant integration time for each pixel, as shown in Figure 1(b). The scanning area is confined to a centered region measuring 30% of the edge length to achieve consistently a high responsivity (see supplementary information). A baseline correction, accounting for the existing lateral responsivity variation of the drum, is made for each line scan based on a one-dimensional median filter (red line in Figure 1(b)). Then, the average drum frequency shift ($\Delta f$) is calculated for each scan point, which is shown in Figure 1(c). The integration time for each scan point is ensured to be larger than the thermal relaxation time constant of ~200 ms for the drum (see supplementary information). From the individual 1-D line scans, the 2-D images are stitched together, as shown in Figure 1(d). A corresponding scanning electron microscopy (SEM) image of the same AuNPs is shown in Figure 1(e). The position of nanoparticles in our scan overlaps perfectly with the corresponding SEM image.

Due to the dielectric nature of silicon nitride, the SEM imaging of the AuNPs on the drum was limited by the charging effect. In contrast, the thermal insulation of the dielectric drum is the basis of our high responsivity, which results in significantly enhanced imaging contrast, as shown in Figure 1(d) and (e). Hence, nanomechanical photothermal microscopy provides a new option for the imaging and analysis of nanoparticles on poor- or non-conducting substrates, which could be challenging for SEM. More scans with corresponding SEM images of AuNPs with different diameter ranging from 30 nm to 200 nm are provided in the supplementary information.

**Analysis of AuNPs absorption cross-section**

Optical probing of plasmonic nanostructures typically yields the sample extinction, which is the sum of scattering and absorption. To show that nanomechanical photothermal microscopy uniquely allows for a pure absorption sample analysis, we studied single AuNPs with varying diameters, as shown in Figure 2(a). The average relative frequency shifts ($\delta f = \Delta f/f_0$) for the individual particles as a function of peak irradiance are plotted in Figure 2(b). From the relative frequency shifts, the absorbed power $P_{abs}$ can be calculated from

$$P_{abs} = \delta f/R, \qquad (1)$$

with R being the relative responsivity of the drum, which was measured each time before sampling (see supplementary information). As expected, the measured relative frequency shifts for AuNPs of a specific dimension scale linearly with the used irradiance I of the laser beam. The absorption cross-section $\sigma_{abs}$ can then be calculated by

$$\sigma_{abs} = P_{abs}/I, \qquad (2)$$

which is represented by the linear fits in Figure 2(b). In Figure 2(c), the extracted average $\sigma_{abs}$ are plotted as a function of particle size. It shows that the $\sigma_{abs}$ values follow the absorption model of the Mie theory with high accuracy, compared to the steeper slope expected for pure scattering. The advantage of detecting absorption instead of scattering becomes obvious from Figure 2(c), with the corresponding scattering cross-section decaying significantly faster for AuNps with dimensions below 200 nm.

As shown in the inset of Figure 2(c), the current laser wavelength of 633 nm does not match the maximal absorption peak of small AuNPs. Switching to a laser with a wavelength closer to 500 nm, for example, would significantly improve the absorption signal of sub-10 nm AuNPs in particular. But even with the non-optimal heating laser wavelength, we have shown the capability for a quantitative absorption analysis and imaging of plasmonic samples.

**Effect of tensile stress on responsivity**

The relative responsivity of a rectangular nanomechanical drum can be described to a good approximation by [27]

$$R \approx \frac{\alpha E}{8\pi\kappa h\sigma}\left(\frac{2-v}{1-v} - 0.642\right), \quad (3)$$

which is a first-order Taylor series approximation of the responsivity of a circular drum. The parameters are the thermal expansion coefficient $\alpha = 2.2 \times 10^{-6}$ K$^{-1}$, Young's modulus $E = 250$ GPa, thermal conductivity $\kappa = 3$ W/(m·K), Poisson's ratio $v = 0.23$, drum thickness $h = 50$ nm, and tensile stress $\sigma$ (more details in supplementary information). Obviously, the frequency response for a unit absorbed power of a drum resonator can be enhanced by reducing the tensile stress $\sigma$. The relationship between tensile stress in the silicon nitride drum and its responsivity are thus investigated both experimentally and theoretically.

To analyze the effects of tensile stress and optimize the responsivity systematically, we used silicon nitride drums with an intrinsic tensile stress of 1 GPa, 250 MPa and 30 MPa.

A drum with 30 MPa was further treated with 50 W oxygen plasma for 10 sec, 20 sec, and 25 sec in order to reduce the tensile stress.[28] The oxygen plasma forms a thin silicon oxynitride layer with a compressive stress on the drum surface. In this way, the tensile stress was further reduced to 6 MPa, 1.2 MPa, and 0.8 MPa, respectively. In Figure 3(a), the experimentally obtained responsivities, based on the specific absorption cross-sections of the measured AuNPs, for all drums with varying stress are plotted and compared with the theoretical model equation (3). The measured responsivities follow the model with high accuracy. Only for the lowest tensile stress values, the model (3) based on the first-order Taylor series approximation (blue dashed line in Figure 3(a)), fails to describe the experimental values, since the thermal stress becomes of the same order of magnitude as the intrinsic tensile stress. Therefore, the full model is plotted (black solid line in Figure 3(a)), which fits the measured values well.

The minimum detectable absorption cross-section in Figure 3(a) was calculated with (2) from the measured responsivities, assuming a general relative frequency stability of $\delta f$ =$10^{-7}$ (see supplementary information), and a peak irradiance of $I = 330$ µW/µm². From the minimum detectable absorption cross section, the detection and imaging capability can be classified in two regimes: (i) single nanoparticle regime with $\sigma_{abs} > 10^{-19}$ m² and (ii) single molecules with $\sigma_{abs} < 10^{-19}$ m². With the current setup, single-nanoparticle resolution could be easily achieved even with a stoichiometric silicon nitride drum with 1 GPa of tensile stress. As presented in Figure 2, drums with 250 MPa of tensile stress allow the detection and analysis of AuNPs down to 30 nm with high reliability. Drums with a tensile stress of 30 MPa are responsive enough to detect and analyze 10 nm AuNPs, as demonstrated in Figure 1(d).

To directly to show the responsivity enhancement of tensile stress reduction, we plot in Figure 3(b) scans of 10 nm AuNPs on a drum with different tensile stress but for the same irradiance. Clearly, contrast increases drastically as the initial tensile stress of 30 MPa is reduced to 6 MPa and 1.2 MPa after the oxygen plasma treatments. From the theoretical absorption cross-section of $\sigma_{abs} = 4.2 \times 10^{-19}$ m² of a single 10 nm AuNP (based on Mie theory for a wavelength of 633 nm) the corresponding responsivities of $R = \delta f \cdot \sigma_{abs} \cdot I =$ 50 kW$^{-1}$ and 311 kW$^{-1}$, can be calculated for the drums with decreased stress from 30 MPa to 6 MPa and 1.2 MPa, respectively.

After the plasma-induced stress reduction, the obtained responsivities all show the potential for single molecule imaging, with expected smallest detectable $\sigma_{abs} < 10^{-19}$ m², as can be seen from Figure 3(a). The limit for stress reduction is given by the required dynamic range, and by background absorption of the probing laser, whose heating induced stress reduction has to be smaller than the intrinsic tensile stress.

**Single molecule imaging and analysis**

In order to demonstrate single-molecule sensitivity in a convincing and reliable fashion, a fluorescence dye was chosen due to the well-established characteristics, such as single-step photobleaching and blinking, which make the unambiguous identification of single molecules possible. Additionally, a fluorescent dye allows for reference imaging with fluorescence microscopy. Atto 633 was adopted in this study due to the matching of the absorption peak (~630 nm) with our available pump laser wavelength. Atto 633 has an absorption peak at 630 nm and a fluorescence peak at 651 nm. For better identification of molecule scans and for reference, we also added highly fluorescent beads during sampling. Figure 4(a) shows a corresponding nanomechanical photothermal scan of one bead, featuring the highest signal, and three single Atto 633 molecules.

Fluorescent molecules dissipate heat either directly via non-radiative relaxation from excited states, or via vibronic relaxation before and after radiative transitions. For Atto 633, the absorption cross section at 633 nm is calculated from the molar extinction coefficient $\varepsilon_{633} = 1.27 \times 10^5$ M$^{-1}$cm$^{-1}$ to be $\sigma_{abs} = 4.84 \times 10^{-20}$ m$^2$. With current pumping irradiance of 330 µW/µm$^2$, this results in a total absorption of 16.7 pW. In Atto 633 only 38.34% of the total absorbed power dissipated into heat through the two non-radiative pathways, which results in a total dissipated power of $P_{abs} = 6.3$ pW. This dissipated power results in a relative frequency detuning of $\delta f = P_{abs} \cdot R = 2.0 \times 10^{-6}$ for a calculated responsivity (based on the non-approximated model shown as solid black line in Figure 3(a)) of $R = 459$ kW$^{-1}$ for a silicon nitride drum with $\sigma = 0.8$ MPa with a resonance frequency of $f_0 = 87$ kHz. This gives an absolute frequency detuning of $\Delta f = 0.25$ Hz, which fits well with the detuning frequencies of $\Delta f = 0.19$ Hz, 0.22 Hz and 0.24 Hz from the single Atto 633 molecules presented in the close-up view in Figure 4(b). The slight differences can be explained by differences in the orientation of the dye molecules resulting in different extent of dipolar excitation by the slightly polarized probing laser.

From the given responsivity, it is now possible to estimate the sensitivity in terms of absorbed power. For an integration time equal to the thermal response time of 200 ms, we measured an Allan deviation of $\Delta f = 4 \times 10^{-3}$ Hz. This gives a relative frequency resolution of $\delta f = 4.6 \times 10^{-8}$, resulting in a sensitivity of 45 fW/Hz$^{1/2}$. For a single Atto 633 molecules this gives a signal-to-noise ratio of 63. An absorption profile of a single Atto 633 molecule is presented in Figure 4(c).

Since all measurements are done in vacuum at $10^{-4}$ mbar, the probability of photobleaching is negligible due to the lack of oxygen. In order to proof our single molecule detection sensitivity, we made reference measurements of the same samples using fluorescence microscopy, as shown in Figure 4(d). The same pattern of three molecules and a fluorescent bead is visible in both, our photothermal scans and the fluorescence image. Similar signals as shown in Figure 4(d) where either found to bleach in a single-step (Figure 4(e)) or exhibit a strong blinking behavior (Figure 4(f)) – two characteristic properties of single molecules.

**Conclusions**

In contrast to shot-noise limited optical probing of sample absorption, nanomechanical probing of the photothermal heating is insensitive to scattering and limited only by thermomechanical noise. The achieved sensitivity at room temperature of 45 fW/Hz$^{1/2}$ is an improvement of five and two orders of magnitude compared to photothermal contrast microscopy,[23,24] and shot-noise limited optical absorption measurements,[21,22] respectively. Furthermore, this mechanical photothermal detection scheme provides an alternative for imaging of nano-object on non-conducting substrates with improved contrast comparing with SEM. Both the obtained signal-to-noise ratio of 63 for a single molecule and image resolution can further be improved by moving from a 50x to a 100x objective. Besides the imaging of single non-fluorescent molecules, our technique allows a unique pure absorption analysis of nanoplasmonic structures, such as AuNPs or nanostructures with more complexity. For that purpose, the response can be significantly improved by shifting from the 633 nm laser wavelength to around 500 nm, where small AuNPs in particular have an absorption peak maximum. By using an infrared scanning laser the presented method can be readily extended in order to perform single-molecule infrared absorption spectroscopy.[27,29,30]

**Methods**

**Fabrication of nanomechanical resonator substrates.** The pre-stress of the silicon nitride thin film could be defined by the chemical composition of silicon nitride from the low-pressure chemical vapor deposition (LPCVD) process and the subsequent $O_2$ plasma treatment.[28] 50-nm-thick LPCVD stoichiometric silicon nitride with pre-stress of around 1 GPa and silicon-rich silicon nitride with pre-stress of around 250 MPa and 30 MPa on Si (100) wafers (Hahn Schickard) were used in present experiments. For drums, a simple bulk micromaching process was done from back side, which etched the window with KOH (40 wt%) with etching rate of 50 μm/hr.[31] Reactive ion etching with $O_2$ plasma with RF power of 50 W was done on 30 MPa drums for 10 sec. and 20 sec., respectively to further reduce the stress to average of 6 MPa and 1.2 MPa. For trampolines, LPCVD stoichiometric silicon nitride with pre-stress of around 1 GPa is used. The silicon nitride trampolines were first defined by front side photolithography and reactive ion etching, and then released by KOH (40 wt%) etching from the back side. $O_2$ plasma treatment was subsequently done with RF power of 150 W for 30 sec. to further reduce the pre-stress to around 200 MPa. All drums used in the work had lateral dimension of 530 μm x 530 μm.

**Sampling of analytes.** Reactant-free gold nanoparticles with diameters of 10, 20, 30, 50, 80, 90, 100, 150 and 200 nm in 0.1 mM PBS stabilized suspension solution (Sigma-Aldrich) were first diluted in micropur deionized water (18 MΩ-m, Milli-Q) with ratio of 1:40 at room temperature, respectively, and syringed through PTFE membrane syringe filters (Acrodisc, Sigma-Aldrich) with a pore size of 200 nm to reduce the aggregations. The filtered solutions were then spin-coated on silicon nitride drums at 2000 rpm for 10 sec. and 4000 rpm for 20 sec. to evenly distribute the nanoparticles. 1 µl of Atto 633 (BioReagent, Sigma-Aldrich) stock solution (1 mg/ml) was diluted in micropur deionized water (Milli-Q) with the ratio of 1:1000 at room temperature, and directly sampled on silicon nitride trampolines with pipette followed by air-drying at room temperature for immediate measurements. All sampling processes were also done in cleanroom. The SEM images of nanoparticles were characterized by Hitachi SU8030 with 3 kV acceleration voltage and 20 pA emission current.

**Measurement electronics.** The real-time optical readout of a laser Doppler vibrometer (MSA-500, Polytec) after a digital velocity decoder was directly captured by a lock-in amplifier (HF2LI, Zurich Instrument) for the tracking of resonance frequency, as shown in figure 1 (b). Five power steps of vibrometer 633 nm laser was used: 305, 170, 68.3, 45.5 and 21.2 µW, and focus by 50× objective (0.55 N.A., Mitutoyo) with fwhm = 0.9 µm and spot size of ~1.53µm. A piezoelectric element (NAC2003, Noliac) was connected to the output of the lock-in amplifier for actuation. A frequency sweep was performed before every measurement and scanning for the phase locking and to optimize actuation voltage. All experiments were done under high vacuum condition with chamber pressure below $10^{-4}$ mbar.

**Finite element simulation of nanomechanical resonator.** The mechanical simulations for responsivity of drum were done by COMSOL with thermostress modul. The vacuum condition was simulated by allowing heat transfer via radiation and conduction only in two dimensions through the anchor. Photothermal heating was simulated by putting a point heat source at center.

**Fluorescence microscopy.** An inverted microscope (Axiovert 200, Zeiss, Germany) equipped with a 100x NA 1.46 oil immersion objective (Plan-Apochromat, Zeiss) and a 640 nm diode laser (iBeam smart, Toptica, Germany) with a power density of 0.5 kW/cm² at the sample was used for imaging. The drum was put upside down on a #1.5 coverslip (24x60 mm, Menzel, Germany) and up to 1000 images were recorded with an illumination time of 5 ms and a delay time of 10 ms. Timing protocols were generated and executed with an in-house written program package implemented in LabVIEW (National Instruments, USA). After appropriate filtering (zt488/640 rpc, Chroma, USA; FF01-538/685-25, Semrock, USA), signals were detected with a back illuminated EMCCD-camera (iXon Ultra 897, Andor, UK). Experiments were carried out at 22°C.

# References

1.  Liang, F., Guo, Y., Hou, S. & Quan, Q. Photonic-plasmonic hybrid single-molecule nanosensor measures the effect of fluorescent labels on DNA-protein dynamics. 1–11 (2017).

2.  Kukura, P. *et al.* High-speed nanoscopic tracking of the position and orientation of a single virus. *Nat. Methods* **6,** 923–927 (2009).

3.  Mashaghi, A. *et al.* Label-free characterization of biomembranes: from structure to dynamics. *Chem. Soc. Rev.* **43,** 887–900 (2014).

4.  Cognet, L. *et al.* Single metallic nanoparticle imaging for protein detection in cells. *PNAS* **100,** (2003).

5.  Dantham, V. R. et al. Label-free detection of single protein using a nanoplasmonic–photonic hybrid microcavity. *Nano Lett.* **13,** 3347–3351 (2013).

6.  Baaske, M. D., Foreman, M. R. & Vollmer, F. Single-molecule nucleic acid interactions monitored on a label-free microcavity biosensor platform. *Nat Nanotechnol* **9,** 933–939 (2014).

7.  Vollmer, F., Arnold, S. & Keng, D. Single virus detection from the reactive shift of a whispering-gallery mode. *Proc. Natl Acad. Sci. USA* **105,** 20701–20704 (2008).

8.  Zijlstra, P., Paulo, P. M. R. & Orrit, M. Optical detection of single non-absorbing molecules using the surface plasmon resonance of a gold nanorod. *Nat Nanotechnol* **7,** 379–382 (2012).

9.  Li, B.-B. *et al.* Single nanoparticle detection using split-mode microcavity Raman lasers. *Proc. Natl Acad. Sci. USA* **111,** 14657–14662 (2014).

10. Zhu, J. *et al.* On-chip single nanoparticle detection and sizing by mode splitting in an ultrahigh-Q microresonator. *Nat. Photonics* **4,** 46–49 (2010).

11. Bohren, C. F. & Huffman, D. R. Absorption and Scattering of Light by Small Particles. *Wiley* 544 (1998). doi:978-0-471-29340-8

12. Boyer, D., Tamarat, P., Maali, A., Lounis, B. & Orrit, M. Photothermal Imagin of Nanometer-Sized Metal Particles Among Scatterers. *Science (80-. ).* **297,** 1160–1163 (2002).

13. L Cognet, S Berciaud, D Lasne, B. L. Photothermal methods for single nonluminescent nano-objects. *Anal. Chem.* **80,** 2288–2294 (2008).

14. Nedosekin, D. A., Galanzha, E. I., Dervishi, E., Biris, A. S. & Zharov, V. P. Super-Resolution Nonlinear Photothermal Microscopy. *Small* **10,** 135–142 (2014).

15. Kim, J., Galanzha, E. I., Shashkov, E. V, Moon, H. & Zharov, V. P. Golden carbon nanotubes as multimodal photoacoustic and photothermal high-contrast molecular agents. *Nat. Nanotechnol.* **4,** 688–694 (2009).

16. Hong, X., Dijk, E. M. P. H. Van, Hall, S. R. & G, B. Background-Free Detection of Single 5 nm Nanoparticles through Interferometric Cross-Polarization Microscopy. *Nano Lett.* **11,** 541–547 (2011).

17. Lindfors, K., Kalkbrenner, T., Stoller, P. & Sandoghdar, V. Detection and Spectroscopy of Gold Nanoparticles Using Supercontinuum White Light b. 3–6 (2004). doi:10.1103/PhysRevLett.93.037401

18. Kukura, P., Celebrano, M., Renn, A. & Sandoghdar, V. Single-Molecule Sensitivity in Optical Absorption at Room Temperature. *J. Phys. Chem. Lett.* **1,** 3323–3327 (2010).

19. Heylman, K. D. *et al.* Optical microresonators as single-particle absorption spectrometers. *Nat. Photonics* **10,** 788–795 (2016).

20. Lo, S. S., Devadas, M. S., Gregory, S. & Hartland, V. Optical detection of single nano-objects by transient absorption microscopy. *Analyst* **138,** 25–31 (2013).

21. Celebrano, M., Kukura, P., Renn, A. & Sandoghdar, V. Single-molecule imaging by optical absorption. *Nat. Photonics* **5,** 95–98 (2011).

22. Chong, S., Min, W. & Xie, X. S. Ground-State Depletion Microscopy: Detection Sensitivity of Single-Molecule Optical Absorption at Room Temperature. *J. Chem. Lett.* **1,** 3316–3322 (2010).

23. Gaiduk, A., Yorulmaz, M., Ruijgrok, P. V & Orrit, M. Room-temperature detection of a single molecule's absorption by photothermal contrast_supplement. *Science (80-. ).* **330,** 353–356 (2010).

24. Gaiduk, A., Ruijgrok, P. V., Yorulmaz, M. & Orrit, M. Detection limits in photothermal microscopy. *Chem. Sci.* **1,** 343 (2010).

25. Schmid, S., Wu, K., Larsen, P. E., Rindzevicius, T. & Boisen, A. Low-power photothermal probing of single plasmonic nanostructures with nanomechanical string resonators. *Nano Lett.* **14,** 2318–2321 (2014).

26. Larsen, T., Schmid, S., Villanueva, L. G. & Boisen, A. Photothermal analysis of individual nanoparticulate samples using micromechanical resonators. *ACS Nano* **7,** 6188–6193 (2013).

27. Kurek, M. *et al.* Nanomechanical Infrared Spectroscopy with Vibrating Filters for Pharmaceutical Analysis. *Angew. Chemie - Int. Ed.* **56,** 3901–3905 (2017).

28. Luhmann, N. *et al.* Effect of oxygen plasma on nanomechanical silicon nitride resonators. 1–5 (2017).

29. Andersen, A. J. *et al.* Nanomechanical IR spectroscopy for fast analysis of liquid-dispersed engineered nanomaterials. *Sensors Actuators, B Chem.* **233,** 667–673 (2016).

30. Yamada, S., Schmid, S., Larsen, T., Hansen, O. & Boisen, A. Photothermal infrared spectroscopy of airborne samples with mechanical string resonators. *Anal. Chem.* **85,** 10531–10535 (2013).

31. Williams, K. R., Member, S., Gupta, K., Member, S. & Wasilik, M. Etch Rates for Micromachining Processing — Part II. **12,** 761–778 (2003).

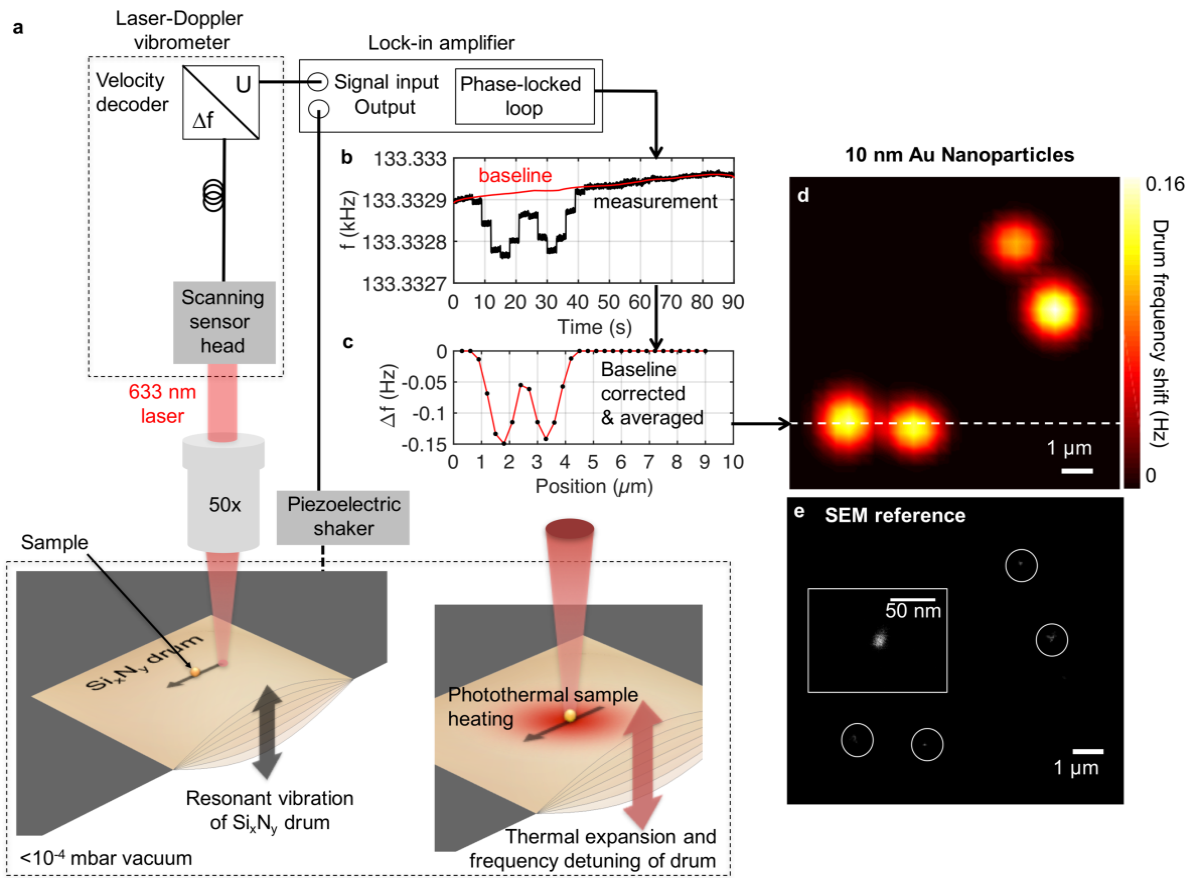

*Figure 1. (a) Schematic description of working principle and detection scheme. (b) PLL tracking of drum resonance frequency (fundamental (1,1) mode) for a line scan over two 10 nm Au particles with the 633 nm scanning laser with a power of 305 µW. The particles are measured on a silicon-rich silicon nitride drum with 30 MPa of tensile stress. (c) Frequency shift of the line scan after baseline correction and individual scan point averaging. (d) Corresponding 2D scan of 10 nm gold nanoparticles. (e) Reference SEM image of AuNPs with a higher magnification image in the inset.*

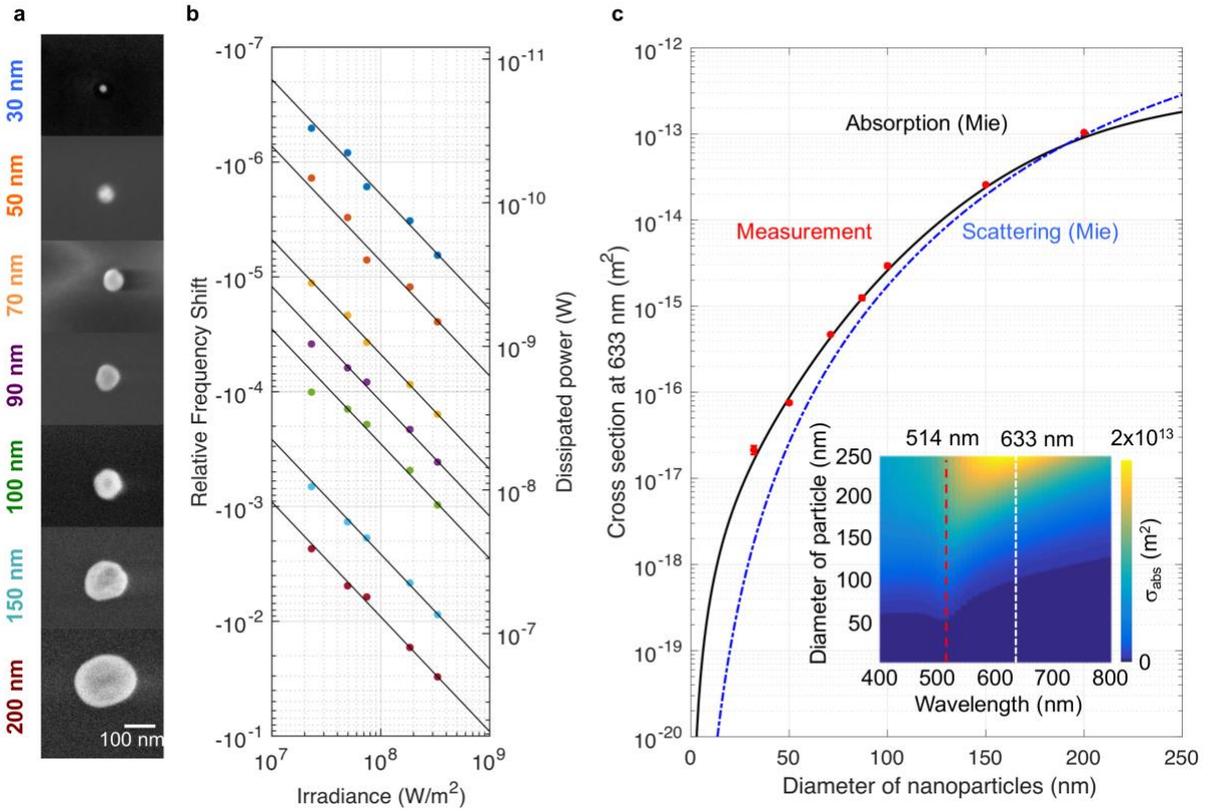

*Figure 2. (a) SEM images of AuNPs with different diameter. (b) Measured relative frequency shift of silicon nitride drums with 250 MPa of tensile stress for AuNPs with different diameter as a function of irradiance. The dissipated power is calculated from the relative frequency shift via (1). (c) Average (>20 particles per size) absorption cross-section as a function of particle size, compared to Mie absorption model and scattering model at 633 nm. Inset: Absorption cross-section spectra for varying AuNP diameter. 514 nm and 633 nm are marked as red and white dashed lines, respectively.*

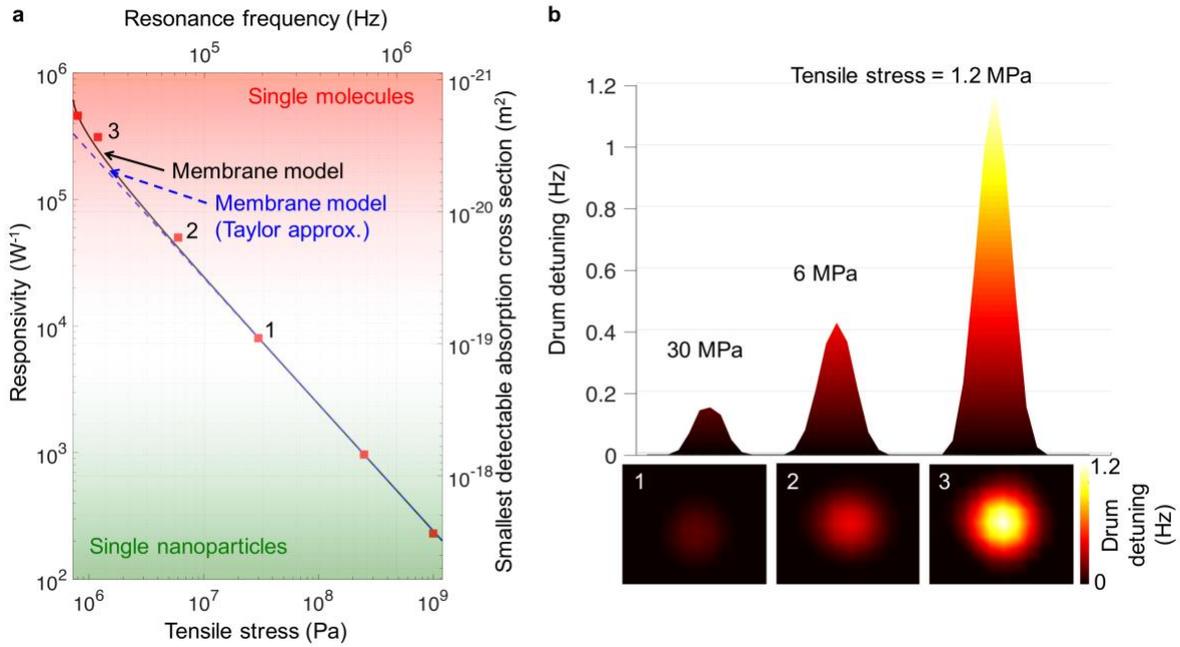

*Figure 3. (a) Measured responsivity of silicon nitride drum with different tensile stress. (red square) Black solid line is the membrane model and blue dashed line is the Taylor-approximated membrane model. The tensile stress of each drum could be measured by the resonance frequency, and the minimum detectable absorption cross section at 633 nm could be derived from responsivity and Allan deviation minimum as noise level. (b) Drum detuning profile and scans of 10 nm AuNPs with different tensile stress of drum. The marking numbers correspond to the measurements in (a).*

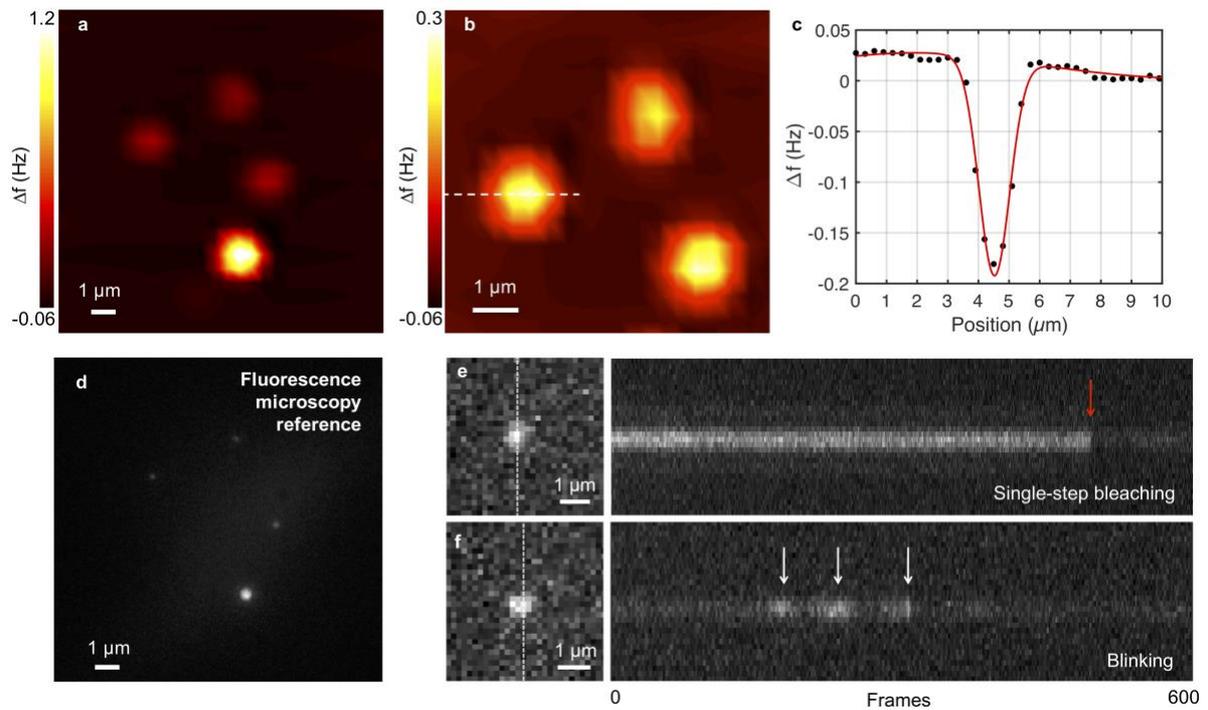

*Figure 4. (a) Scan of three single Atto 633 molecules and one fluorescent bead, measured with the (4,4) mode of a silicon nitride drum with a tensile stress of 0.8 MPa. (b) Close-up of the three single molecules. (c) Profile cut through the absorption peak of a single Atto 633 molecule. (d) Reference fluorescence microscopy image (a). (e) Single-step bleaching (red arrorw) and (f) blinking (white arrow) of Atto 633 measured with fluorescent microscopy.*

# Supplementary Information:
## Single-molecule optical absorption imaging by nanomechanical photothermal sensing at room temperature

Miao-Hsuan Chien[1], Mario Brameshuber[2], Gerhard J. Schütz[2], and Silvan Schmid[1]

[1] Institute of Sensor and Actuator Systems, TU Wien, Gusshausstrasse 27-29, 1040 Vienna, Austria.
[2] Institute of Applied Physics, TU Wien, Wiedner Hauptstrasse 8-10, 1040 Vienna, Austria.

## Characterization of drum resonators

To calibrate and characterize the mechanical properties of the membrane resonator for NMSA measurements, a pristine silicon-rich silicon nitride membrane with lateral dimensions of 530 µm × 530 µm and a thickness of 50 nm was first scanned with the vibrometer laser of different power steps ranging from 305 to 21.2 µW focused through a 50× objective. To minimize the heat dissipation via gas convection and increase the quality factor of the resonator, all experiments were done at $10^{-4}$ mbar vacuum. A linear fit between the frequency and corresponding power is done to subtract the frequency of the fundamental mode, as shown in Figure S1. The fundamental mode (1, 1) resonance frequency ($f_0$) of membrane is measured to be ~389 kHz. The tensile stress of the membrane could be extracted from the eigenfrequency model for a membrane[1]

$$f_0 = \pi \left(\frac{\sqrt{n^2+j^2}}{L}\right)^2 \sqrt{\frac{-2D_p + \frac{\sigma h L^2}{\pi^2}}{2\rho h}},$$

where with length of membrane $L$, mode numbers ($n$, $j$) to be (1, 1), and mass density $\rho$ of 3000 kg/m³, thickness h of 50 nm and the $D_p$ is the flexural rigidity of membrane given by

$$D_p = \frac{Eh^3}{12(1-\nu^3)},$$

For tensile stress higher than ~100 Pa regime, effect of flexural rigidity could be neglected and the model could be reduced to

$$f_0 = \frac{\sqrt{n^2+j^2}}{2L}\sqrt{\frac{\sigma}{\rho}}.$$

The tensile stress of membrane is calculated to be ~250 MPa. The quality factor of the membranes is ~1 million. The thermal time constant of the membranes is measured to be less than 200 ms. A theoretical model for the membrane responsivity $R$ could be deducted from equation of motion for stressed membrane and Fourier's law of heat transfer, and a membrane model has already been established in a previous study[2]:

$$R = \frac{1-\sqrt{1-\frac{\alpha EP}{4\pi\kappa h\sigma}(\frac{2-v}{1-v}-\frac{1}{1.56})}}{P},$$

with first order Taylor approximation, such that

$$R = \frac{\alpha E}{8\pi\kappa h\sigma}\frac{1.3584-0.35844v}{1-v},$$

where the thermal expansion coefficient of silicon nitride $\alpha$ = 2.2 ppm /K, the Young's modulus $E$ = 250 GPa, the thermal conductivity $\kappa$ = 3 W/m-K, the tensile stress $\sigma$ = 250 MPa, and the poisson's ratio $v$ = 0.23. A comparison of used constant in previous studies is provided in Table S1. Both finite element simulation and the theoretical model predict a frequency responsivity of ~980 W$^{-1}$, which fits the measured value with absorption of silicon nitride membrane within 0.5%. The frequency shift results from this background absorption of membrane are subtracted in every measurement of nanoparticles and molecules. As the tensile stress of silicon nitride drum reduces to the same order with the thermal stress induced by the background absorption of the laser power, the approximation started to deviate.

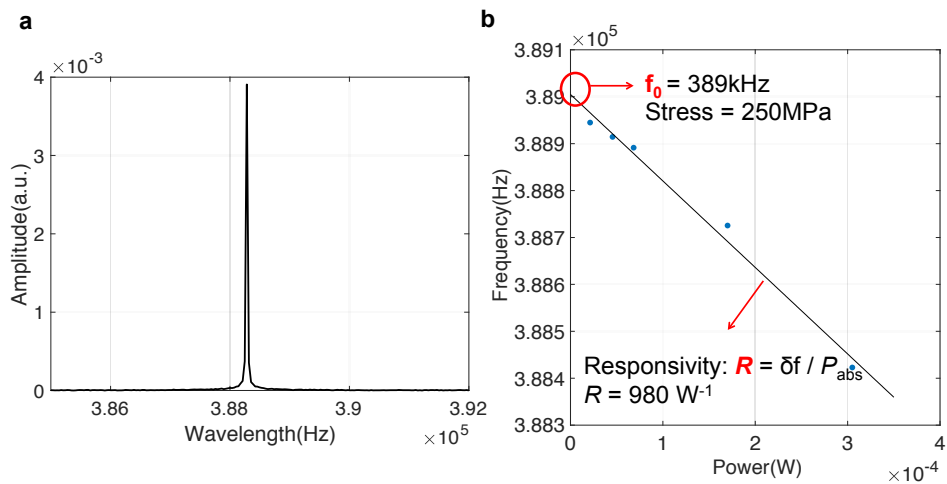

Figure S1. (a) Measured resonance frequency for the pristine 250 MPa silicon-rich silicon nitride membrane with the dimension of 530 μm × 530 μm × 50 nm (L × L × h) under power step of 305 μW. The average quality factor is in the $10^6$ regime at a vacuum of $10^{-4}$ mbar. (b) Measurement and subtraction of stress (intercept) and responsivity (slope) for the same membrane by switching the power steps of the LDV. The average absorption of the silicon nitride drum is 0.5%.

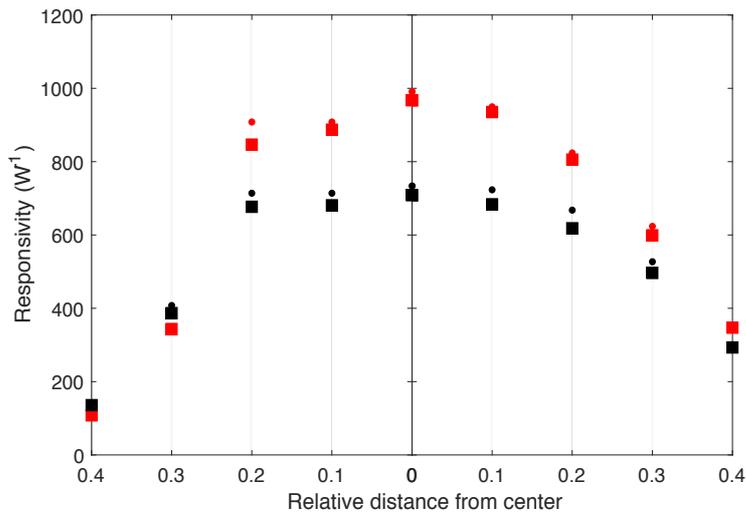

Figure S2. Responsivity with respect to different positions of the silicon nitride drum. Right side of the figure goes in (110) direction of the silicon window while the left side goes in (111) direction. The square notation is the measured values, and the dotted notation is the value from COMSOL simulations by assuming a point heat source on the silicon nitride resonator and allowing only two dimensional heat transfer. The NMSA measurements were only done in the center 30% of the drum to ensure the high and consistent responsivity.

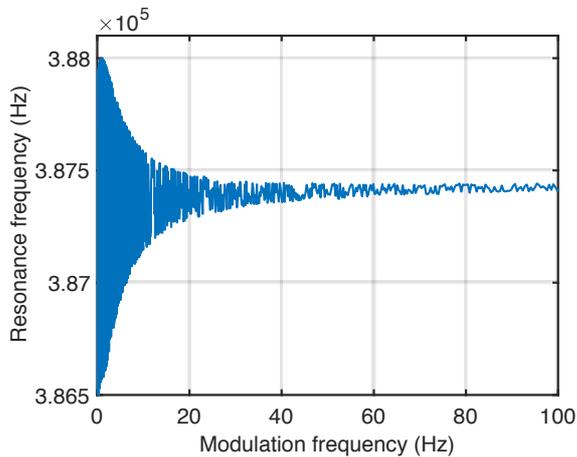

Figure S3. Measurements of the time constant of the silicon nitride drum. A second laser was focused on the center of the drum and modulated with frequency ranges from 0.1 Hz to 100 Hz, and the resulting detuning of the drum resonance frequency was monitored. The time constant was measured to be less than 200 ms.

| | Stress | Thermal conductivity $\kappa$ (W/(m-K)) | Thermal expansion coefficient $\alpha$ ($K^{-1}$) | Young's modulus $E$ (GPa) | Poisson's ratio $\upsilon$ | Specific heat compacity (J/(kg-K)) |
|---|---|---|---|---|---|---|
| Present study | 0.8 MPa-1GPa | 3 | 2.2 ppm | 250 | 0.23 | 700 |
| Y. Toivola et al. J. Appl. Phys. 2013[3] | Silicon-rich (Regardless of stress) | N. A. | 2.2 ppm | 300 | N. A. | N. A. |
| T. Larsen et al. ACSnano 2013[4] | 170 MPa | 20 | 1 ppm | 250 | 0.23 | 700 |
| S. Schmid et al. Nano Lett. 2014[5] | 900 MPa | 2.5 | 1.23 ppm | 240 | N. A. | N. A. |
| S. Yamada et al. Anal. Chem. 2013[6] | 200 MPa | 2.5 | 1.23 ppm | 250 | N. A. | N. A. |
| H. Ftouni et al. Phy. Rev. Lett. B 2015[7] | Regardless of stress | 2.5-3 (at 300 K) | N. A. | N. A. | N. A. | 800 (at 300 K) |

Table S1. Comparisons of constants for silicon nitride.

## Frequency resolution of membrane resonators

The integration time was optimized by means of the Allan deviation with minimum frequency fluctuations for an integration time of 400 ms. The Allan deviation, which is measured to be in the range of $8\times10^{-8}$ to $10^{-7}$ for membranes with 250 MPa tensile stress, is limited for short integration times by white noise ($\tau^{-0.5}$) and for long integration times by drift, as shown in Figure S4.

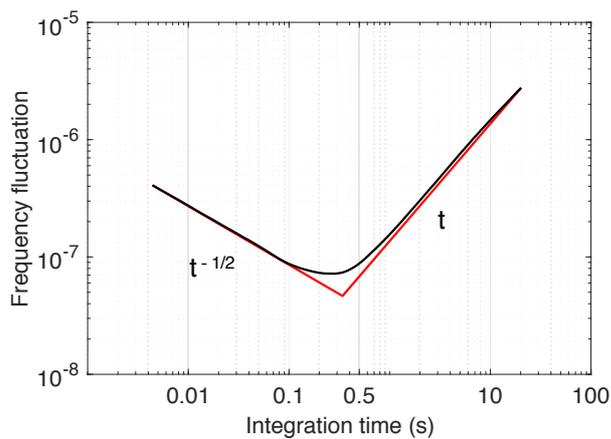

Figure S4. Measured Allan deviation for the 250 MPa silicon-rich silicon nitride membrane with the dimensions of 530 μm × 530 μm × 50 nm (L × L × h).

## Imaging and characterization of AuNPs

NMSA and the corresponding SEM images of single nanoparticles with diameters of 100 nm, 50 nm and 30 nm are shown in Figure S5. Since the contrast is created by the relative frequency shift resulting from the absorption of the nanoparticles, AuNPs with different dimensions show clearly different response intensity. The signal from 100 nm one is roughly an order of magnitude higher than the 50 nm one and two order of magnitude to the 30 nm one under same scanning irradiance I using the same membranes. NMSA imaging of AuNPs with diameters of 100 nm, 50 nm, and 30 nm and their corresponding SEM images are shown in Figure S5. The resulting maximum relative frequency shifts for the different AuNPs of ~$1.5\times10^{-3}$ for 100 nm AuNPs, ~$1.5\times10^{-4}$ for 50 nm AuNPs and ~$1.5\times10^{-5}$ for 30 nm AuNPs, correspond to absorption cross sections of ~$2\times10^{-15}$ m$^2$, ~$2\times10^{-16}$ m$^2$, and ~$2\times10^{-17}$ m$^2$, respectively. The histogram of absorption cross section distribution for AuNPs with different dimensions are shown in Figure S6.

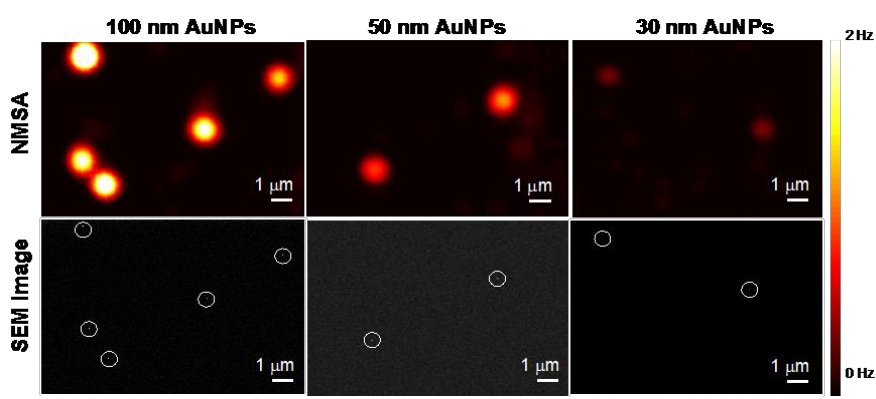

Figure S5. NMSA imaging and the corresponding SEM images of AuNPs with diameter of 100 nm, 50 nm and 30 nm.

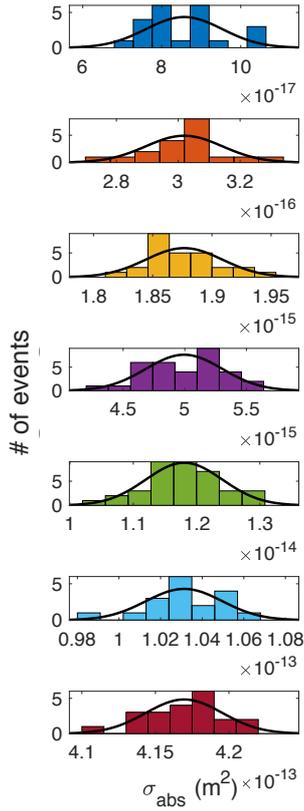

Figure S6. The histograms of absorption cross section distribution of AuNPs with different diameters. The mean value and standard deviation were extracted and all histograms show near-Gaussian distribution, consistent with the datasheet provided by the company.

## AuNPs absorption cross section calculations and simulations

The absorption cross section $\sigma_{abs}$ of individual Au nanoparticles was calculated based on Mie theory absorption model,

$$\sigma_{abs} = \frac{2\pi}{\lambda} Im\{\alpha\}$$

with the consideration of retardation effects and corrected polarizability $\alpha$ of[13]

$$\alpha = 3V\varepsilon_m \frac{1 - 0.1(\varepsilon + \varepsilon_m)\theta^2/4}{\frac{\varepsilon + 2\varepsilon_m}{\varepsilon - \varepsilon_m} - \frac{(0.1\varepsilon + \varepsilon_m)\theta^2}{4} - \frac{2\varepsilon_m^{\frac{3}{2}}\theta^3}{3}}$$

where $\theta = 2\pi r/\lambda$, electrical permittivity $\varepsilon_m$ for vacuum is 1, r is the radius of the nanoparticle, and the electrical permittivity $\varepsilon$ is taken from Johnson and Christy's report.[14] To evaluate the substrate effect of Silicon nitride drum, finite difference time domain (FDTD) simulations were done by putting a single AuNP on 50 nm-thick silicon nitride or without any substrate. The results of 100 nm AuNP are shown in Figure S7, which shows similar spectrum for cases with or without silicon nitride substrates. The differences of $\sigma_{abs}$ at 633 nm is within one order of magnitude. And the results of FDTD simulations without substrates are also quite close with Mie model (Figure S8). The simulated value is slightly higher, which may result from the meshing artifacts or the monitor positions.

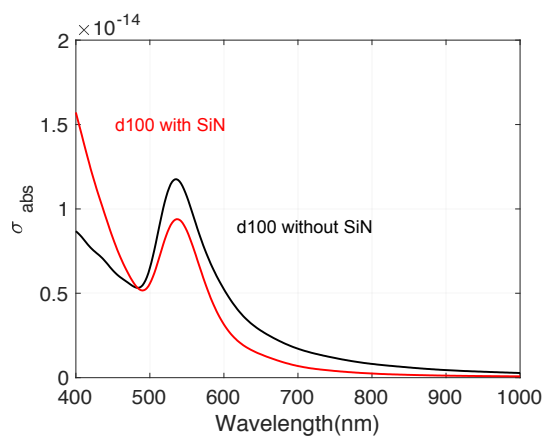

Figure S7. The FDTD simulated absorption cross section of AuNP with diameter of 100 nm with and without 50 nm silicon nitride membrane underneath.

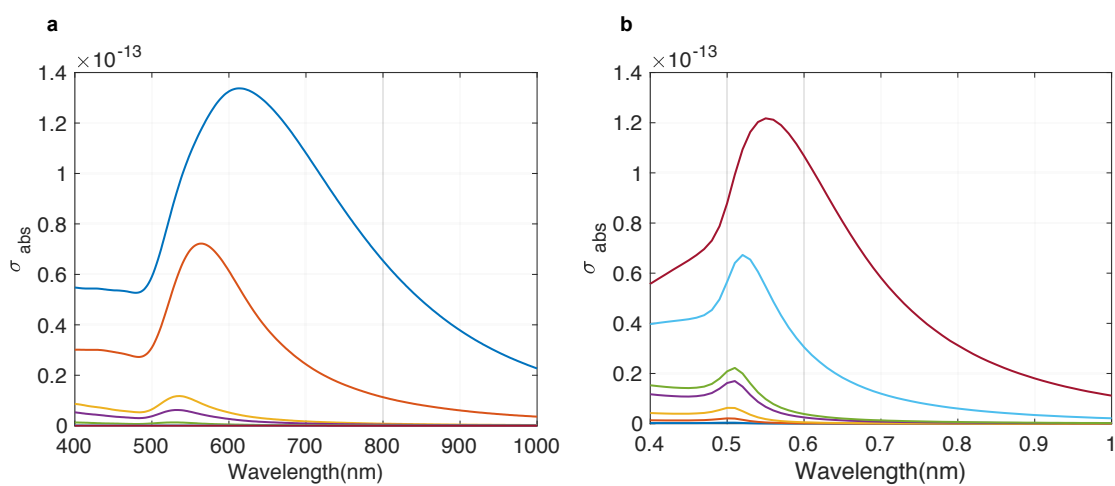

Figure S8. (a) The FDTD simulated absorption cross section for AuNPs with diameter of 200 nm, 150 nm, 100 nm, 90 nm, 70 nm, 50 nm and 30 nm. (b) The calculated absorption cross section for AuNPs with diameter of 200 nm, 150 nm, 100 nm, 90 nm, 70 nm, 50 nm and 30 nm from the Mie theory with the consideration of retardation effect.

## Characterization of Atto 633 single molecules

The fluorescence properties of Atto 633 could be found on the official site of AttoTec. The important characteristics that are related to present study are listed in Table S2, and the extinction spectrum is plotted in Figure S9, with maximum absorbance at 630 nm. The absorbance for 633 nm excitation is 97.6%, resulting in molar extinction coefficient $\varepsilon_{633}$ of $1.27 \times 10^5$ $M^{-1}$ $cm^{-1}$. The absorption cross section at 633 nm could thus by calculated by

$$\sigma_{633} = \varepsilon_{633} \times 10^3 \times \ln(10)/N_A$$

in units of cm$^2$, with the Avogadro constant $N_A$ of 6.02×10$^{23}$, with corresponding absorption power of 16.658 fW. And the fraction of heat dissipation $\eta_{diss}$ could be calculated by

$$\eta_{diss} = (1 - \eta_{fl}) + \frac{\eta_{fl}(\frac{1}{\lambda_{exc}} - \frac{1}{\lambda_{fl}})}{\frac{1}{\lambda_{exc}}}$$

with fluorescence quantum yield $\eta_{fl}$, excitation wavelength $\lambda_{exc}$ and fluorescence wavelength $\lambda_{fl}$. The heat dissipation was calculated to be 6.263 fW, with the silicon nitride drum responsivity of 458645 W$^{-1}$ and resonance frequency of 87 kHz, resulting in frequency detuning of ~0.244 Hz for single molecules.

| | $\tau_F$ (ns) | $\eta_{fl}$ | $\eta_{heat}$ | $\varepsilon_{max}$ (M$^{-1}$ cm$^{-1}$) | $\varepsilon_{633}$ (M$^{-1}$ cm$^{-1}$) | $\sigma_{633}$ (m$^2$) | $\lambda_{exc}$ (nm) | $\lambda_{fl}$ (nm) |
|---|---|---|---|---|---|---|---|---|
| Atto 633 | 3.2 | 64% | 38.34% | 1.3×10$^5$ | 1.27×10$^5$ | 4.84×10$^{-20}$ | 633 | 651 |

Table S2. Properties of Atto 633 with excitation laser of 633 nm.

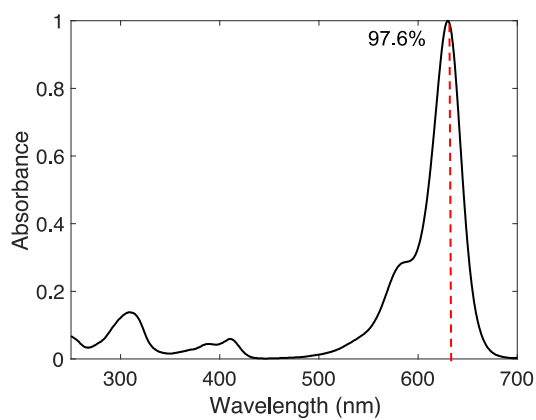

Figure S9. Extinction spectrum of Atto 633 provided by AttoTec. The absorbance at 633 nm was indicated to be 97.6%.

## Frequency resolution used for sensitivity calculation

The Allan deviation of the drum used for single molecule imaging is shown in Figure S10.

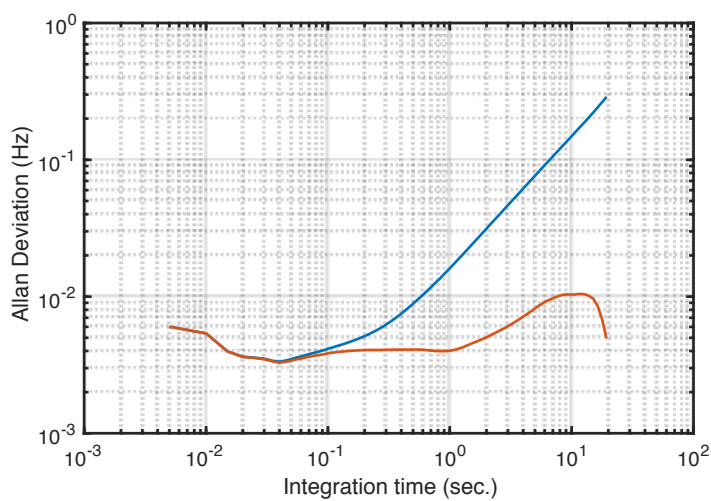

Figure S10. Measured Allan deviation for uncorrected (blue line) and baseline corrected (orange line) frequency data, for the 0.8 MPa silicon-rich silicon nitride membrane with the dimension of 530 μm × 530 μm × 50 nm (L × L × h) used for single molecule imaging.

# References

1. Schmid, S., Villanueva, L. G. & Roukes, M. L. *Fundamentals of Nanomechanical Resonators*. (2016).
2. Kurek, M. *et al.* Nanomechanical Infrared Spectroscopy with Vibrating Filters for Pharmaceutical Analysis. *Angew. Chemie - Int. Ed.* **56,** 3901–3905 (2017).
3. Toivola, Y., Thurn, J., Cook, R. F., Cibuzar, G. & Roberts, K. Influence of deposition conditions on mechanical properties of low-pressure chemical vapor deposited low-stress silicon nitride films. *J. Appl. Phys.* **94,** 6915–6922 (2003).
4. Larsen, T., Schmid, S., Villanueva, L. G. & Boisen, A. Photothermal analysis of individual nanoparticulate samples using micromechanical resonators. *ACS Nano* **7,** 6188–6193 (2013).
5. Schmid, S., Wu, K., Larsen, P. E., Rindzevicius, T. & Boisen, A. Low-power photothermal probing of single plasmonic nanostructures with nanomechanical string resonators. *Nano Lett.* **14,** 2318–2321 (2014).
6. Yamada, S., Schmid, S., Larsen, T., Hansen, O. & Boisen, A. Photothermal infrared spectroscopy of airborne samples with mechanical string resonators. *Anal. Chem.* **85,** 10531–10535 (2013).
7. Ftouni, H. *et al.* Thermal conductivity of silicon nitride membranes is not sensitive to stress. *Phys. Rev. B - Condens. Matter Mater. Phys.* **92,** 1–7 (2015).

8. Magnes, J. *et al.* Quantitative and Qualitative Study of Gaussian Beam Visualization Techniques. 1–5 (2006).

9. Andersen, A. J. *et al.* Nanomechanical IR spectroscopy for fast analysis of liquid-dispersed engineered nanomaterials. *Sensors Actuators, B Chem.* **233,** 667–673 (2016).

10. Schmid, S., Kurek, M., Adolphsen, J. Q. & Boisen, A. Real-time single airborne nanoparticle detection with nanomechanical resonant filter-fiber. *Sci. Rep.* **3,** 1288 (2013).

11. Ekinci, K. L., Yang, Y. T. & Roukes, M. L. Ultimate limits to inertial mass sensing based upon nanoelectromechanical systems. *J. Appl. Phys.* **95,** 2682–2689 (2004).

12. Sansa, M. *et al.* Frequency fluctuations in silicon nanoresonators. *Nat. Nanotechnol.* **11,** 552–558 (2016).

13. Myroshnychenko, V. *et al.* Modelling the optical response of gold nanoparticles. *Chem. Soc. Rev.* **37,** 1792–1805 (2008).

14. Johnson, P. B. & Christy, R. W. optical contstants of noble metals.pdf. *Phys. Rev. B* **6,** 4370–4379 (1972).